\documentclass{osa-article}

\journal{osajournal}

\articletype{Research Article}

\usepackage{textcomp}
\usepackage{amsmath}    
\usepackage{verbatim}   
\usepackage{color}      
\usepackage{subfigure}  
\usepackage{placeins}	
\usepackage{epstopdf}
\usepackage{xspace}
\usepackage{hyperref}
\usepackage{makecell}
\usepackage[none]{hyphenat}
\hypersetup{
  colorlinks=true,
  linkcolor=blue,
  citecolor=blue,
  urlcolor=blue,
  breaklinks=true}
 
\usepackage{units}   
\usepackage[separate-uncertainty=true,detect-weight=true, detect-family=true]{siunitx}    
  
\def\fm#1{\ifmmode #1 \else $#1$\fi}
\newcommand{\Itw}{\fm{I_{2\omega}}\xspace}
\newcommand{\Iw}{\fm{I_{\omega}}\xspace}
\newcommand{\ktw}{\fm{k_{2\omega}}\xspace}
\newcommand{\kw}{\fm{k_{\omega}}\xspace}
\newcommand{\deff}{\fm{d_\mathrm{eff}}\xspace}

\newcommand{\Dk}{\fm{\Delta k}\xspace}

 \newcommand{\uproman}[1]{\uppercase\expandafter{\romannumeral#1}}
 \newcommand{\al}{$^{27}\mathrm{Al}^+$\xspace}
 \newcommand{\hg}{$^{199}\mathrm{Hg}^+$\xspace}
 \newcommand{\ind}{$^{115}\mathrm{In}^+$\xspace}

 \DeclareSIUnit{\rad}{rad}
  \DeclareSIUnit{\mrad}{mrad}

\usepackage{lineno}

\begin{document}

\title{Phase-stabilized UV light at \SI{267}{\nano\m} through twofold second harmonic generation}

\author{B. Kraus\authormark{1,2}, F. Dawel\authormark{1}, S. Hannig\authormark{1,2}, J. Kramer\authormark{1}, C. Nauk\authormark{1}, P. O. Schmidt, \authormark{1,2,3,*}}

\address{\authormark{1}Physikalisch-Technische Bundesanstalt, Bundesallee 100, 38116 Braunschweig, Germany\\
\authormark{2}Deutsches Zentrum f{\"u}r Luft- und Raumfahrt e.V. (DLR), Institut für Satellitengeod{\"a}sie und Inertialsensorik, c/o Leibniz Universit{\"a}t Hannover, DLR-SI, Callinstraße 30b, 30167 Hannover, Germany\\
\authormark{3}Institut f{\"u}r Quantenoptik, Leibniz Universit{\"a}t Hannover, Welfengarten 1, 30167 Hannover, Germany}

\email{\authormark{*}Piet.Schmidt@ptb.de}

\begin{abstract}Providing phase stable laser light is important to extend the interrogation time of optical clocks towards many seconds and thus achieve small statistical uncertainties. We report a laser system providing more than \SI{50}{\micro\W} phase-stabilized UV light at \SI{267.4}{\nano\m} for an aluminium ion optical clock. The light is generated by frequency-quadrupling a fibre laser at \SI{1069.6}{\nano\m} in two cascaded non-linear crystals, both in single-pass configuration. In the first stage, a \SI{10}{\milli\m} long PPLN waveguide crystal converts \SI{1}{\W} fundamental light to more than \SI{0.2}{\W} at \SI{534.8}{\nano\m}. In the following \SI{50}{\milli\m} long DKDP crystal, more than \SI{50}{\micro\W} of light at \SI{267.4}{\nano\m} are generated. An upper limit for the passive short-term phase stability has been measured by a beat-node measurement with an existing phase-stabilized quadrupling system employing the same source laser. The resulting fractional frequency instability of less than \SI{5e-17} after \SI{1}{s} supports lifetime-limited probing of the \al clock transition, given a sufficiently stable laser source. A further improved stability of the fourth harmonic light is expected through interferometric path length stabilisation of the pump light by back-reflecting it through the entire setup and correcting for frequency deviations. The in-loop error signal indicates an electronically limited instability of $1\times 10^{-18}$ at \SI{1}{s}.
\end{abstract}

\section{Introduction}

\label{sec:intro}
Many applications in quantum physics require phase stable light sources. 
Optical clocks stand out for their stringent requirement on phase stability to achieve long interrogation times and thus small statistical uncertainties \cite{ludlow_optical_2015}. 
This is particularly important for single or few ion optical clocks, owing to their limited signal-to-noise ratio, requiring lifetime-limited interrogation times to achieve competitive statistical uncertainties \cite{leroux_-line_2017,peik_laser_2006,riis_optimum_2004,itano_quantum_1993}. Starting from lasers stabilized to ultra-stable cavities \cite{alnis_subhertz_2008,ludlow_compact_2007,nazarova_vibration-insensitive_2006,young_visible_1999, dawkins_ultra-stable_2009,didier_946-nm_2019,dube_narrow_2009,hafner_8_2015,hafner_transportable_2020,jiang_making_2011,kessler_sub-40-mhz-linewidth_2012,leibrandt_spherical_2011,millo_ultrastable_2009,robinson_crystalline_2019,thorpe_measurement_2010,vogt_demonstration_2011,webster_force-insensitive_2011,webster_thermal-noise-limited_2008,zhang_ultrastable_2017} that can provide coherence times exceeding \SI{11}{\s} \cite{matei_15_2017}, several schemes have been devised to even further extend the coherence time of the clock interrogation light by employing multiple ensembles of clocks \cite{borregaard_efficient_2013,hume_probing_2016,kessler_heisenberg-limited_2014,rosenband_exponential_2013, takamoto_frequency_2011, schioppo_ultrastable_2017, dorscher_dynamical_2020}. To maintain the high phase coherence, the effect of disturbances on the probe light due to fluctuations of the refractive index caused by temperature or pressure variations within the optical medium (air or optical fibers) on the way towards the clock atoms must be eliminated. This is typically achieved through a Michelson interferometer, where a short reference arm is compared in a heterodyne measurement with the frequency-shifted laser radiation that has traveled through the medium and back again, with its frequency corrected by an acousto-optical frequency shifter \cite{bergquist_frontiers_1992, ma_delivering_1994}.
However, several optical clock transitions are in the ultraviolet (UV) spectral range. Examples include the \al ion quantum logic optical clock with a clock transition wavelength of
\SI{267}{\nano\m} \cite{rosenband_frequency_2008,brewer_27+_2019,clements_lifetime-limited_2020,chou_frequency_2010,leibrandt_trapped-ion_2017, si-jia_chao_observation_2019, rosenband_observation_2007,leopardi_measurement_2021}, the \hg ion clock with a clock transition at \SI{282}{\nano\m} \cite{diddams_optical_2001,oskay_single-atom_2006,rosenband_frequency_2008}, the \ind ion clock operated at a wavelength of \SI{237}{\nano\m} \cite{ohtsubo_frequency_2017,ohtsubo_frequency_2020,wang_absolute_2007}, and the neutral atom mercury lattice clock at \SI{266}{\nano\m} \cite{mcferran_neutral_2012,ohmae_direct_2020,takamoto_frequency_2015,yamanaka_frequency_2015}.
The UV laser light in this wavelength regime is typically generated through fourth harmonic generation (FHG) of IR lasers. A common method of FHG generation is given by twofold SHG in non-linear media often using pump-beam power enhancement in a cavity for high efficiency \cite{ko_high-power_2015, lo_all-solid-state_2014, wang_high-power_2012, hume_two-species_2010, hannig_highly_2018, schmitt_single_2009, manzoor_high-power_2022, shaw_stable_2021, burkley_highly_2019, kaneda_continuous-wave_2009, kaneda_continuous-wave_2008, scheid_750_2007, herskind_second-harmonic_2007, sakuma_generation_2004, freegarde_design_2001, zimmermann_all_1995}. The challenge is to transfer the phase stability from the cavity-stabilized IR laser through all optical paths including the frequency doubling stages to the UV light. 
It has been shown that highly efficient single-pass waveguide doublers, converting IR radiation into the visible wavelength regime, do not introduce additional noise at a level of $10^{-19}$\cite{herbers_phase_2019, delehaye_residual_2017}. However, waveguide doublers for conversion into the UV are not available. Frequency doubling systems employing a pump beam resonant enhancement cavity are expected to add negligible noise, since the optical path length of the pump beam within the cavity is stabilized by locking the cavity length to the pump beam wavelength \cite{abdel-hafiz_guidelines_2019}. Therefore, phase stabilization is required before, after and between the SHG units, including possible active elements, such as power amplifiers, which significantly increases the complexity of the setup. 

Here we describe and characterize the generation of UV probe light for the \al clock at \SI{267.4}{\nano\m} based on two cascaded single-pass SHG stages, significantly reducing the complexity of the setup and enabling continuous length stabilization of the optical path through back reflection of the IR light at the output of the UV doubling stage. Assuming that the intermediate frequency doubled and the target quadrupled light are common path and dispersion effects are small, we expect that phase stabilisation of the IR light path also provides phase stability of the UV light.
We characterize the performance of the system in terms of output power and phase stability. The latter is quantified by a comparison with a conventional FHG setup with cascaded waveguide and cavity-enhanced SHG units, employing piecewise phase stabilisation. 

\section{Crystal selection}
\label{sec:crystalselection}
According to SHG theory, efficient harmonic generation requires a large nonlinear coefficient \deff, phase matching between the fundamental (pump) and second harmonic field, and a good spatial overlap between these two fields. In the non-depleted pump regime and assuming plane wave interaction, the generated SHG intensity scales quadratically with the pump intensity $\Iw$ \cite{boyd_second-harmonic_1965}  
\begin{equation}
\Itw\propto \deff^2 l^2 \Iw^2 \omega^2\frac{\sin^2\left(\Dk l/2\right)}{\left(\Dk l/2\right)^2},
\end{equation}
where $l$ is the length of the crystal, $\omega$ is the pump frequency and $\Dk=\ktw-2\kw$ is the difference between the wavenumbers of the pump and SHG fields. For strong conversion (pump depletion regime), the relation between SHG and pump intensity becomes linear. For phase-matched fields ($\Dk=0$), the generated SHG in each part of the crystal along the propagation direction is coherently added to the existing SHG field. There are several possibilities to achieve this, depending on the properties of the nonlinear crystal. 
Taking advantage of the birefringence of most nonlinear crystals, critical phase matching can be achieved through a proper choice of polarisation and angle of the incident fields with respect to the crystal axes. However, this results in the SHG field being emitted under an angle with respect to the pump field (walk-off), which reduces the spatial overlap between the fields and therefore SHG efficiency. Some nonlinear crystals offer a temperature-dependence of their refractive indices that enables phase matching within a certain wavelength range (non-critical, temperature, or $90^\circ$ phase matching). In this case, the fields propagate along one of the crystal axis, avoiding walk-off. Finally, quasi-phase matching (QPM) enables an effective $\Dk=0$ through a periodic modulation of the sign of \deff along the propagation direction. The change in sign ensures that the generated SHG field continues to coherently add to the existing SHG field after running out of phase after a coherence length $l_c=\frac{\lambda}{4(n_{2\omega}-n_\omega)}$, where $\lambda$ is the fundamental wavelength and $n_{\omega}$ and $n_{2\omega}$ are the refractive indices of the fundamental and SHG field, respectively. This results in a \deff smaller by a factor of $2/\pi$ compared to non-critical phase matching.
Boyd and Kleinmann have extended the SHG treatment to Gaussian beams \cite{boyd_parametric_1968}, where absorption, divergence and walk-off limit the optimal crystal length and focus inside the crystal. The detrimental effects of Gaussian beam divergence can be overcome by employing waveguide crystals \cite{parameswaran_observation_2002, iwai_high-power_2003, suntsov_watt-level_2021, cho_power_2021, fedorova_efficient_2014, sun_466_2012, sun_efficient_2011}, in which fundamental and harmonic fields are guided in a single transverse mode structure. Cascaded single-pass SHG with mode overlap between the fundamental and harmonic beams requires either QPM or non-critical phase matching, significantly limiting the crystal choices. We have therefore selected a periodically-poled lithium niobate (PPLN) ridge waveguide crystal, manufactured by HC Photonics Corp., for the first doubling stage from \SI{1069.6}{\nano\m} to \SI{534.8}{\nano\m}. The crystal has a length of \SI{10.3}{\milli\meter}, is anti-reflection coated for both wavelengths, and achieves an intrinsic efficiency of \SI{\sim 125}{\percent\per\watt} at a phase-matching temperature of $45(20)$\SI{}{\degreeCelsius}, according to the manufacturer.
The temperature of the QPM SHG crystal needs to be stabilized, since a change in temperature changes the refractive index and -- through thermal expansion -- the periodic poling period. For a waveguide the effective refractive index also depends on the geometry of the waveguide, which is not known to us. We therefore use the expected temperature tuning bandwidth $\Delta$T=\SI{2.8}{\kelvin\centi\meter} 
of a bulk PPLN crystal\cite{fejer_quasi-phase-matched_1992} as a reference.\\ 
For the second doubling stage from \SI{534.8}{\nano\meter} to \SI{267.4}{\nano\meter}, we have chosen a \SI{50}{\milli\meter} long deuterated potassium dihydrogen phosphate (DKDP) crystal, which allows non-critical phase matching at an expected temperature of around \SI{100}{\degreeCelsius}\cite{nikogosyan_nonlinear_2005} and has a non-linear coefficient $\deff\approx$ \SI{0.43}{\pico\meter\per\volt} \cite{noauthor_snlo_nodate} 
The temperature tuning bandwidth of this crystal is given by \SI{3.2}{\kelvin\centi\meter} \cite{noauthor_snlo_nodate}. 
The optimal focus waist of \SI{32}{\micro\meter} \cite{boyd_parametric_1968} of the pump beam inside the crystal results in a calculated single-pass conversion coefficient of \SI{0.15}{\percent\per\watt}. In Tab. \ref{t1}, the key parameters for both crystals are listed. 

\begin{table}[h!]
\caption{Key crystal parameters.
	}
\begin{tabular}{{|c|c|c|}}
\hline
 \textbf{Value}  \ & \  \textbf{PPLN} \ & \  \textbf{DKDP} \  \\
 \hline
 Length in \SI{}{\milli\meter}  & 10.8 & 50 \\
 \hline
 Facet dimensions in \SI{}{\square\milli\meter}  & 0.5\,x\,1.5 & 5.0\,x\,5.0 \\
 \hline
 Transparency range bulk in \SI{}{\nano\meter} & 300\dots 5000 & 200\dots 2100  \\
 \hline
 \makecell{Absorption coefficient pump\\ light in \SI{}{\per\centi\meter}}  & $\approx 0.003$ & $\approx$ 0.004\dots 0.005  \\
 \hline
 \makecell{Absorption coefficient SHG\\ in \SI{}{\per\centi\meter}}  & $\approx 0.03$ & $\approx$ 0.004\dots 0.005  \\
 \hline
 \deff in \SI{}{\pico\meter\per\volt} &  14\dots 16 & $\approx$ 0.43 \\
 \hline
 Conversion coefficient in \SI{}{\percent\per\watt} \ & 125 & 0.15 \\
 \hline
 Hygroscopicity  & none & high \\
 \hline
Expected phase matching temperature in \SI{}{\degreeCelsius}  & $45(20)$ & $\approx$100 \\ 
\hline
 Temp. tuning bandw. in \SI{}{\kelvin\centi\meter}  & 2.8 & 3.2 \\ 
 \hline
 Manufacturer  & HCP & Altechna \\ 
 \hline
\end{tabular}

\label{t1}
\end{table}

\section{Technical description of the setup}

\begin{figure}[ht!]
		\includegraphics[width=7.5cm,height=10cm]{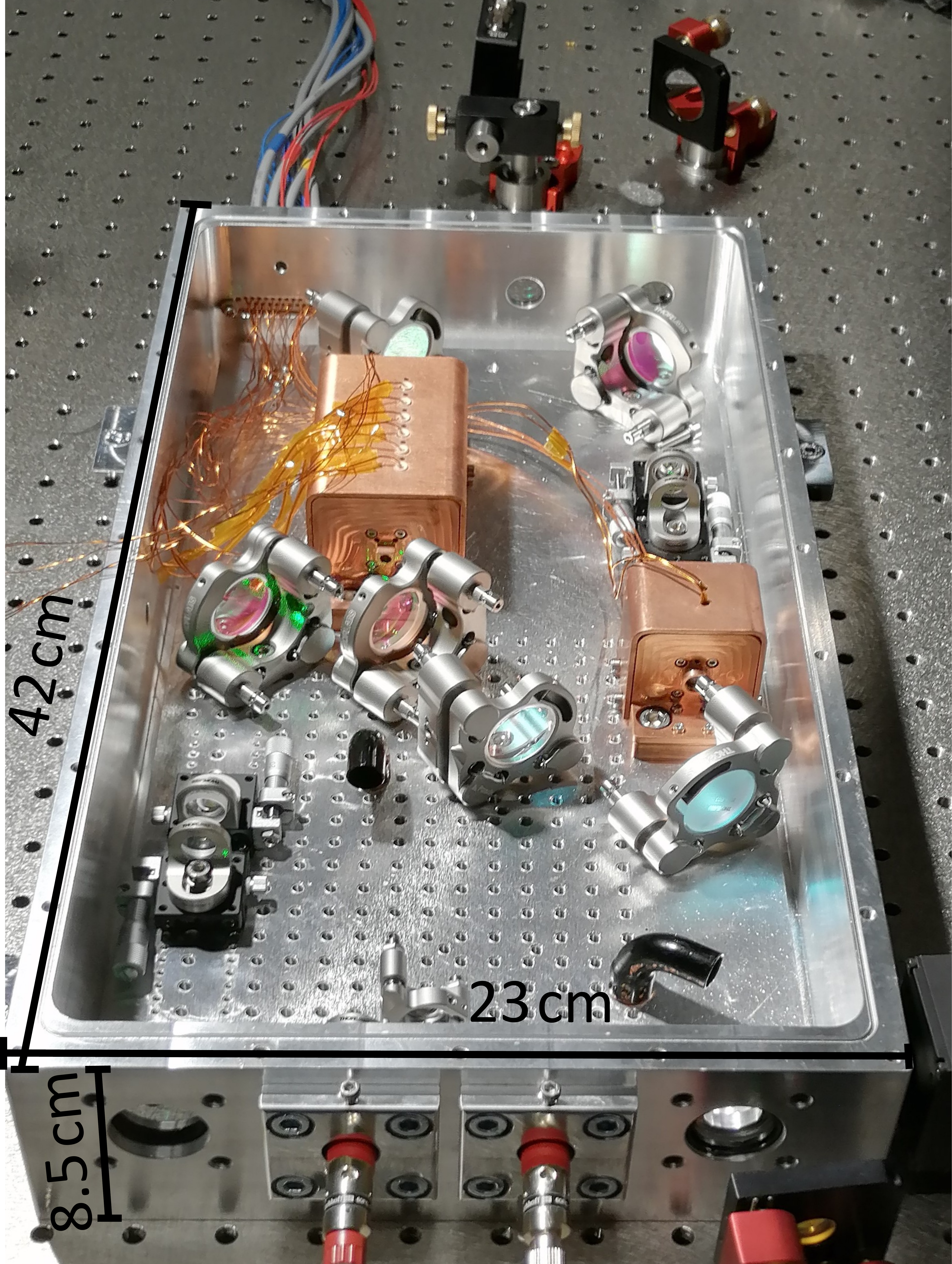}
	\caption{Image of the setup including two cascaded non-linear crystals inside custom ovens covered by copper heat shields.
	}
	\label{Bild1}	
\end{figure}
The FHG setup consists of two cascaded SHG stages. Each stage is placed inside a self-built oven for temperature control. The system is fibre coupled and a phase stabilisation scheme is included to compensate phase noise throughout most of the optical paths.

\subsection{Mechanical design}

 The setup is build on a breadboard inside a hermetically sealed box of size of \SI{42}{\centi\m} x \SI{23}{\centi\m} x \SI{8.5}{\centi\m}, both made from aluminium (see Fig. \ref{Bild1}), which makes it rigid and stable against external perturbations such as temperature, and pressure fluctuations and vibrations. Its compact design allows a high degree of transportability and the integration into a commercial 19 inch rack. The box is sealed with a Viton\textsuperscript{\textregistered} ring and evacuation to avoid air fluctuations is possible via a flange on the lid. Alternatively, damaging of optical components due to UV light can be suppressed by purging the box with clean gas \cite{schroder_laser-induced_2013} for example oxygen or nitrogen. Optical access to the volume is provided by a number of windows which are sealed with indium. 
 The fibre collimators for the two aforementioned optical fibers are mounted to the box from the outside in front of such windows. For focusing of the laser beam into the first SHG stage and recollimation after the first SHG stage non adjustable lenses are integrated in the first oven. For focusing the laser beam into the second stage with a specific beam waist, a telescope consisting of a concave and a convex lens is used. For recollimation an identical telescope in an inverted arrangement is used. All four lenses are mounted on individual linear translation stages to enable a precise adjustment along the direction of beam propagation. The optical power at all wavelengths can be monitored with photodiodes. the photodiodes are placed behind windows outside the box to avoid contamination from out-gassing of components of the photodiodes. All optical components like mirrors and (dichroic) beam-splitters are situated in either adjustable or fixed highly-stable mounts.\\
  \begin{figure}[ht]
 	
 	\includegraphics[width=9cm]{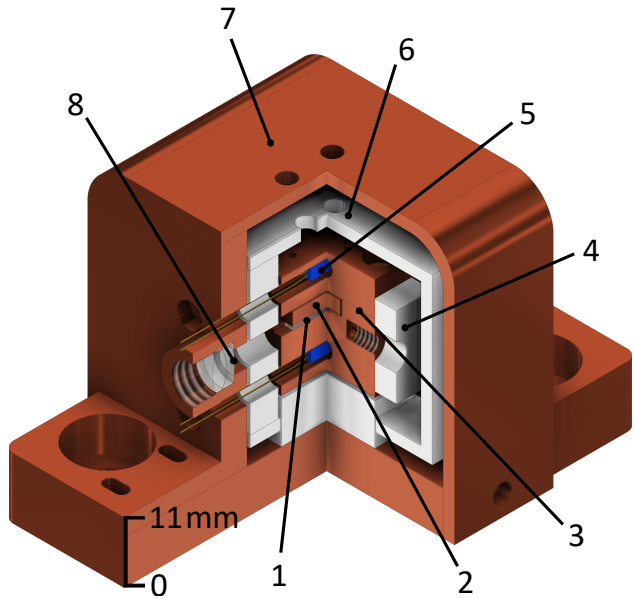}
 	\caption{Cut-away drawing of the oven for the PPLN crystal. 1: waveguide, 2: parts of the socket, 3: fixation parts, 4: Peltier elements, 5: PT1000 temperature sensor, 6: inner MACOR\textsuperscript{\textregistered} shield, 7: outer copper shield, 8. lenses.
 	}
 	\label{ofen}
 \end{figure}
 
\paragraph{Design of the ovens:}  The ovens of both crystals share a similar design and provide an accurate, constant and homogenous phase matching temperature of up to \SI{100}{\degreeCelsius}. A cut-away drawing of the first oven is shown in FIG.~\ref{ofen}. In the first oven the waveguide (1) is placed inside a copper socket (2). To prevent mechanical stress and to ensure thermal contact, the waveguide is wrapped in a single \SI{100}{\micro\m} layer of indium foil. A spring-loaded copper plate is used to push the waveguide against the socket (3), while mechanical tension is avoided. Two Peltier elements (4) are used for heating. The temperature is measured by two PT1000 temperature sensors (5), one of which is integrated in the top part and the other in the bottom part of the socket. A PID controller is used to actively stabilise the oven temperature with a resolution of \SI{0.001}{\degreeCelsius} using one of the two PT1000 temperature sensors. The other PT1000 temperature sensor is used as a monitoring sensor. The connection of the PT1000 temperature sensors and the Peltier elements with the PID controller (Meerstetter TEC-1091) are realized through a vacuum-compatible sealed d-sub multi-pin feed-through, which is integrated into the wall of the box. For thermal isolation of the oven the socket is surrounded by an inner shield made of MACOR\textsuperscript{\textregistered} (6). The remaining temperature difference between the oven and the environment can lead to convection of the surrounding gas. The latter can result in optical power instabilities due to beam pointing fluctuations and phase distortions of the beam. This is suppressed by adding a second shield (7) to the oven. This outer shield is made of copper and is in thermal contact with the breadboard via the copper ground plate of the oven. The apertures for the pump laser beam and SHG beam in the outer shield are realized within an extension of the outer copper shield to further reduce fluctuation in the gas of the pump light path. The outer shield does not exceed \SI{25}{\degreeCelsius} during normal oven operation.
Two lenses (8) for coupling the IR light into the waveguide are directly integrated into the extrusion of the outer shield of the first oven.\\
The DKDB crystal is placed inside a similar oven. However, with \SI{50}{\milli\m} it is longer and thus the spatial homogeneity of the temperature is more critical. Therefore, three rows consisting of three Peltier elements and six PT1000 sensors, three in the bottom part of the socket and three on the top part of the socket are employed. The two outer rows of Peltier elements operate in series on a fixed current. The current of the middle row Peltier elements is controlled by a PID controller for active temperature stabilisation using one of the PT1000 temperature sensor. The other PT1000 temperature sensors can additionally be used for separately monitoring the temperature inhomogeneity.\\
Different from the first oven, no lenses for mode matching are needed inside the second oven as the telescope before the oven serves that purpose. The performance of both ovens at their operational temperature is measured and shown in Tab. \ref{t2}. This includes the long term temperature stability, the maximum heating or cooling ramp (limited for protection of the crystals), the temperature settling time (for temperature stability of < \SI{0.05}{\degreeCelsius}) and the thermalization timescale ($1/e$ time constant of an exponential approach).
\begin{table}[ht]
\caption{Key parameters of the ovens for the nonlinear crystals.
 	}
\begin{tabular}{{|c|c|c|}}
\hline
 Measured value  \ &\ PPLN-oven \ & \ DKDP-oven \  \\
 \hline
Operational temp.  & \SI{44}{\degreeCelsius} & \SI{76}{\degreeCelsius} \\
 \hline
 Long term temp. stability  & \SI{0.01}{\degreeCelsius} & \SI{0.005}{\degreeCelsius}  \\
 \hline
 Max. temp. ramp  & \SI{1}{\degreeCelsius\per\min} & \SI{1}{\degreeCelsius\per\min} \\
 \hline
\ Temp. settling time \ & \SI{30}{\s} & \SI{60}{\s} \\
 \hline
 Thermalization timescale  & \SI{280\pm30}{\s} & \SI{610\pm30}{\s}\\ 
 \hline
\end{tabular}
 	\label{t2}
\end{table}
 
 \subsection{Optical setup}

\begin{figure*}

\includegraphics[width=13.5cm]{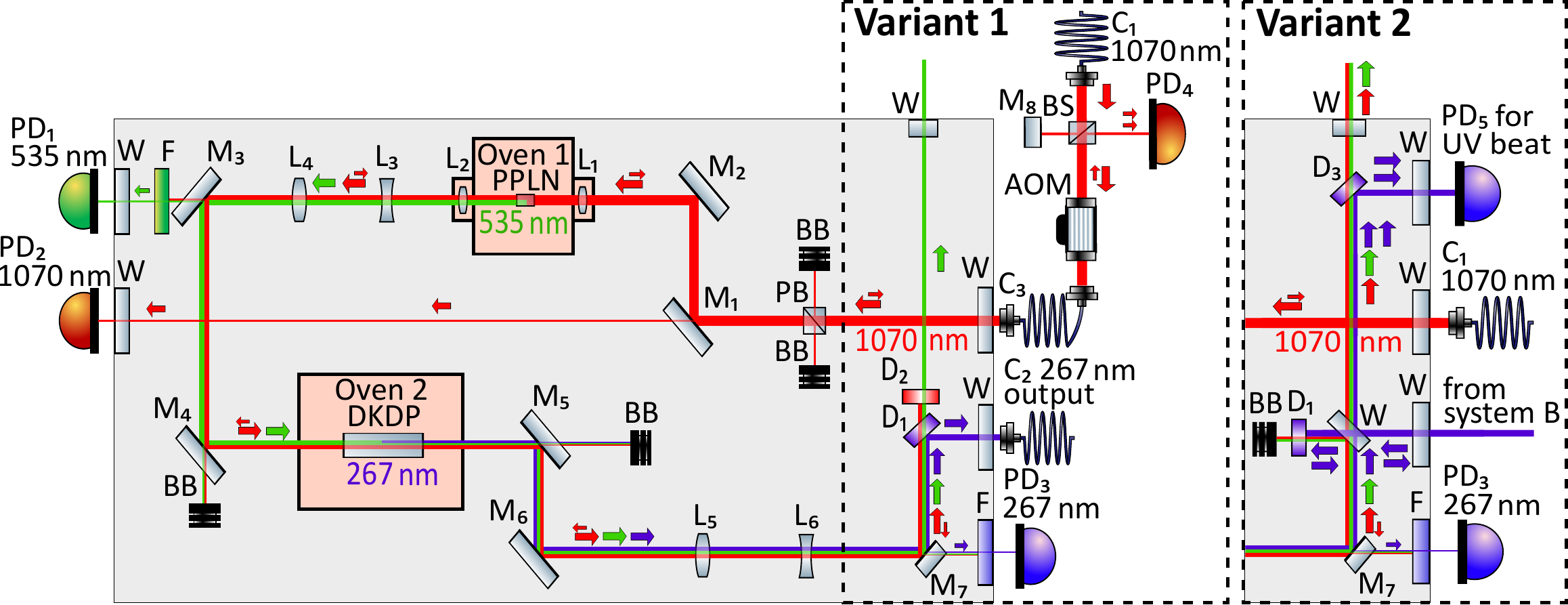}
\caption{Schematic drawing of the setup. An overview of the used parts and the propagation direction of the laser beam is shown. Variant 1: The setup is used as a length-stabilised UV light source. Variant 2: The setup is expanded for a frequency instability measurement of the produced UV light in combination with a second UV light source. C\textsubscript{1}: Collimator, IR light from Laser; C\textsubscript{2}: Collimator, UV light output; PB: Polarizing beam splitter; M: Mirror; L: Lens; BB: Beam block; PPLN: PPLN waveguide; DKDP: DKDP crystal; D: Dichroic mirror; PD: Photo diode; F: Filter; W: Window; BS Beam splitter; AOM: Acousto-optic modulator.}
 	\label{drawing}
 \end{figure*}

 The optical set up is schematically depicted in FIG.~\ref{drawing}. In our case the maximum input power is \SI{1}{\W} at \SI{1069.6}{\nano\m}. The IR light is delivered via a polarisation maintaining fibre to the first collimator outside the box. An AR-coated optical window transmits the light to the inside, where a polarizing beam splitter is used to clean the pump light polarization, resulting in a horizontal polarised pump light beam.

\paragraph{First SHG stage from \SI{1069.6}{\nano\m} to \SI{534.8}{\nano\m}:}
Two mirrors M\textsubscript{1} and M\textsubscript{2} are used to couple the beam into the PPLN waveguide. Mirror M\textsubscript{1} is transparent at the backside allowing the transmitted light to be detected on a photodiode P\textsubscript{1} outside the box for IR optical power monitoring. Lens L\textsubscript{1} has a focal length $f=+\SI{12.5}{\milli\m}$ and is used to match the spatial mode of the pump light at \SI{1069.6}{\nano\m} to the waveguide with an NA of 0.16 in vertical and 0.14 in horizontal 
direction. The PPLN-ridge waveguide is placed inside the first oven.\\ 
It efficiently converts the IR laser light at \SI{1069.6}{\nano\m} into green light at \SI{534.8}{\nano\m} via SHG when actively temperature stabilized at the phase matching temperature $T_{\mathrm{pm}}\approx 44^\circ\textrm{C}$ using type I phase matching. At \SI{534.8}{\nano\m} the NA of the waveguide is specified to be 0.8 in vertical and 0.7 in horizontal direction. The green light is horizontal polarized as two extraordinary photons are converted into one extraordinary with doubled frequency. The outgoing green and IR light beams are collimated with an achromatic AR-coated lens L\textsubscript{2} with $f=+\SI{12.5}{\milli\m}$, resulting in a diameter of \SI{4}{\milli\m} for IR light and \SI{2}{\milli\m} for green light. \\
 \label{sec2}
\paragraph{Second SHG stage \textbf{\SI{535}{\nano\m}} to \SI{267}{\nano\m}:}

For high SHG efficiency of the second stage, the pump beam is focused in the centre of the crystal along its propagation direction. By solving the Boyd-Kleinman expression we find an optimal beam waist size of $w_0=\SI{32}{\micro\m}$ (radius at 1/e\textsuperscript{2} intensity level of the Gaussian beam at the focus point) for our given nonlinear crystal and \SI{200}{\milli\watt} green pump light power.
These calculated optimum focus position and waist for the green pump light are set by a telescope directly behind the first SHG. The telescope consists of two lenses L\textsubscript{3} with $f=-\SI{48}{\milli\m}$ and L\textsubscript{4} with $f=+\SI{75}{\milli\m}$, which are placed on individual linear translation stages.  
The beam positioning between the first and second SHG stage can be adjusted with the two fold mirrors M\textsubscript{3} and M\textsubscript{4}. The first mirror is backside-polished. Therefore, the pump light at \SI{535}{\nano\m} for the second SHG stage can be measured by the photodiode P\textsubscript{2} placed outside the box.\\ 
The green light at \SI{534.8}{\nano\m} is converted into \SI{267.4}{\nano\m} in a \SI{50}{\milli\m} long DKDP crystal using type I phase matching. The crystal is AR coated for both wavelengths and is placed inside the oven, which ensures a homogeneous phase matching temperature along the crystal. The green light is still horizontal polarized, which corresponds to the ordinary polarization in relation to the crystal orientation. The generated UV light is vertical polarized, as two ordinary photons are converted into one extraordinary with doubled frequency.\\
After the DKDP crystal the beam is collimated again with an adjustable two-lens system with L\textsubscript{5} and L\textsubscript{6} on individual linear translation stages and the position is adjusted by two mirrors (M\textsubscript{5,6}). The generated UV light power level is monitored by photodiode P\textsubscript{3} after transmission through a backside-polished mirror M\textsubscript{7} and a subsequent short pass optical filter to remove any residual green and IR light.
In Fig.~\ref{drawing} two variants of the FHG setup are depicted. In variant 1 the setup with a fibre coupled UV output is used as a UV laser source. Here, the reflected light off M\textsubscript{7} is split by a dichroic mirror D\textsubscript{1} in its green and the IR light contribution in pass direction and its UV light contribution in reflection. The UV light leaves the box through a window and can e.g. be coupled into a fibre. Variant~2 shows the setup when it is used as a reverence for second UV laser source or when the frequency stability of the system is measured with a second UV laser source (see \ref{UVlref}).\\

\paragraph{IR light path length stabilization:} The setup is designed for minimal phase noise of the produced UV light. The air sealed box minimises the phase noise contribution due to pressure and temperature fluctuations. For the temperature dependence of the air at the free-space optical path in the box, one can roughly calculate a phase shift of \SI{13}{\rad\per\kelvin}\cite{edlen_refractive_1966} (\SI{1.3}{\rad\per\kelvin} for the IR light path, \SI{3.6}{\rad\per\kelvin} for green light path and \SI{8.9}{\rad\per\kelvin} for UV light path). For the pressure dependence the phase shift is \SI{4}{\rad\per\hecto\pascal}\cite{edlen_refractive_1966} (\SI{0.4}{\rad\per\hecto\pascal} for the IR light path, \SI{1.1}{\rad\per\hecto\pascal} for green light path and \SI{2.6}{\rad\per\hecto\pascal} for UV light path). We do expect effective temperature changes inside the box of less than \SI{10}{\milli\kelvin} per second and pressure changes of less than \SI{0.1}{\pascal} per second. The resulting phase stability is \SI{0.13}{\rad\per\s} for temperature changes and \SI{0.004}{\rad\per\s} for pressure changes. The corresponding fractional frequency change $\gamma(t)$ can be calculated through
\begin{equation}
\gamma(t)=\frac{\Delta\phi}{2\pi t f},    
\end{equation}
where $f$ is the frequency of the light, resulting in
$\gamma(1\,s)\approx$\SI{1.9e-17}{} for temperature changes and $\gamma(1\,s)\approx$\SI{6e-19}{} for pressure changes. On time scales of a few seconds, this poses no limitation to the frequency instability. However, for longer integration times phase stabilisation is necessary. In this setup an interferometric length stabilisation is realised with the IR light and includes both SHGs and the major part of the optical path of the setup. As the beams with all three wavelengths are spatially overlapping until they are separated by D\textsubscript{1}, we expect the beams to undergo similar phase and frequency disturbances. Therefore, the length stabilisation of the IR light path serves as a reference for the UV light for all common length changes. Taking the dispersion of the air into account, the remaining temperature-induced phase shift for UV
light is approximately \SI{0.8}{\mrad\per\kelvin}, while the pressure-induced phase shift for UV
light is approximately \SI{0.2}{\mrad\per\hecto\pascal}\cite{edlen_refractive_1966}. Assuming again temperature changes of \SI{10}{\milli\kelvin\per\s} or \SI{0.1}{\milli\kelvin\per\pascal}, the remaining fractional frequency instability due to the temperature or pressure change in the air is \SI{1e-18} or \SI{3e-20} and thus no limitation for the anticipated clock operation.
The IR length stabilisation is implemented according to the scheme shown in Fig. \ref{drawing}, variant 1. Therefore, the IR light is backreflected by a short pass dichromatic mirror D\textsubscript{2}, closely behind D\textsubscript{1}. The green light is transmitted by D\textsubscript{2} and can be used outside the box through a window. The IR light takes the same path back to the first collimator C\textsubscript{1} and is coupled back into the fibre. The IR light phase stabilisation path consists of an AOM to correct any frequency changes, a beamsplitter, a retro reflective mirror (M\textsubscript{8}) for the reference arm and a photodiode P\textsubscript{4} to measure the heterodyne beat signal between the reference and the signal arm of the interferometer. A phase locked loop between the measured beat signal and a reference frequency is closed by feeding the rf correction signal into the AOM.

\section{Characterisation of the FHG UV light}

 \begin{figure}
 	
 	\includegraphics[width=8.5cm]{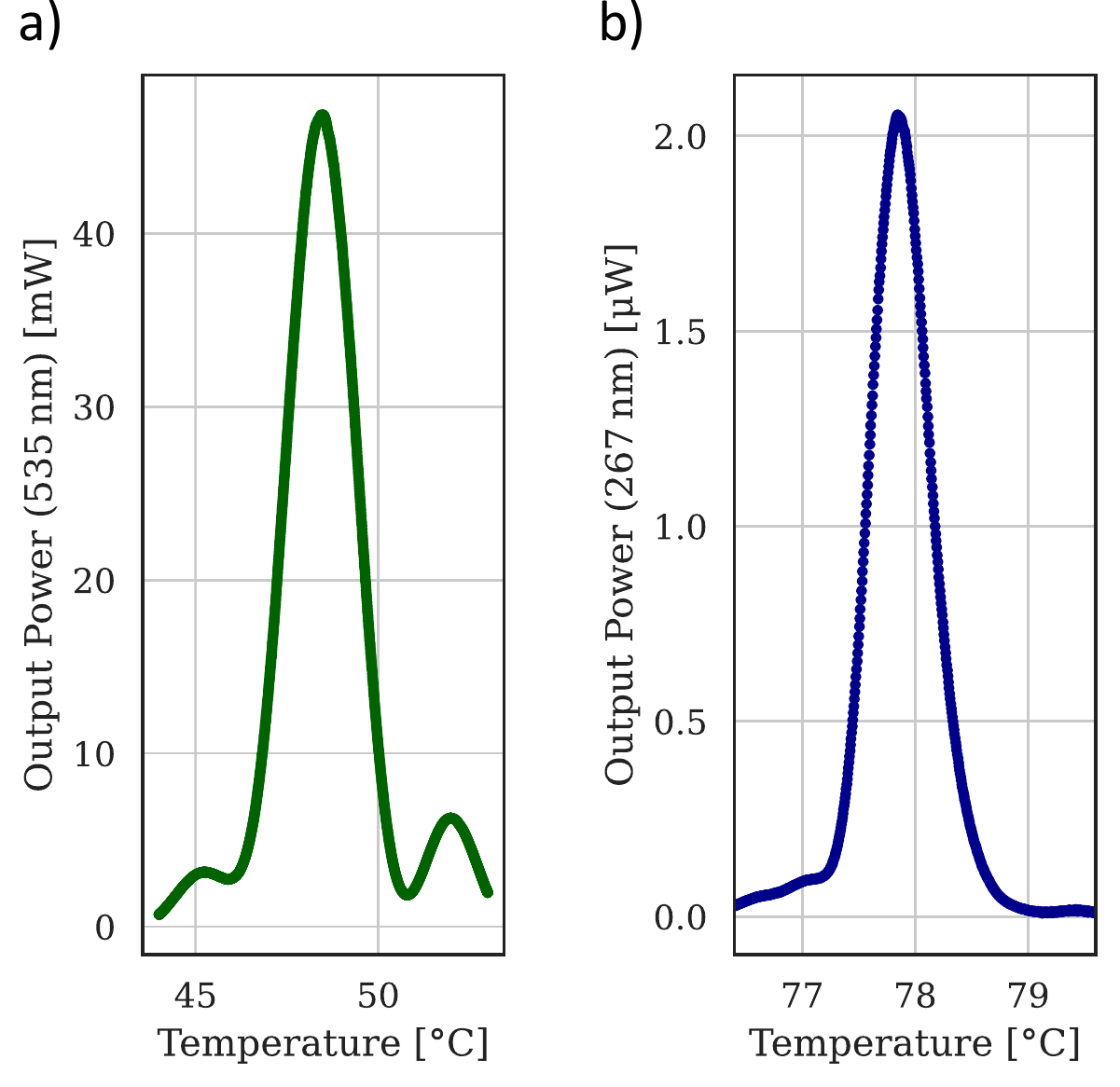}
 	\caption{SHG output power as a function of the oven Temperature a) for the first SHG at \SI{534.8}{\nano\meter} and b) for the second SHG at \SI{267.4}{\nano\meter}. Measured for an IR pump light power of \SI{450}{\milli\W} at \SI{1069.6}{\nano\m}.
 	}
 	\label{tempcur}
 \end{figure}

\subsection{Non-critical phase matching of the SHGs}

Both SHG stages use non-critical phase matching which makes any converted output power sensitve to temperature variations of the nonlinear crystal. The temperature dependence on the output power is depicted in Fig. \ref{tempcur} for the first SHG stage (left) and the second SHG stage (right). For the first SHG the maximum green light output is given at an oven temperature of \SI{48.5}{\degreeCelsius} with a full width half maximum (FWHM) of \SI{2.2}{\degreeCelsius}. The theoretical value for a \SI{10.3}{\milli\m} long bulk crystal is \SI{2.7}{\degreeCelsius} (see section~\ref{sec:crystalselection}). The smaller measured temperature tuning bandwidth compared to the theoretically expected value for a bulk crystal might arise from a different effective reflective index in the waveguide. The maximum output power for the second SHG stage is obtained at an oven temperature of \SI{77.84}{\degreeCelsius} with a FWHM of \SI{0.61}{\degreeCelsius}, which is in good agreement with the theoretically expected value of \SI{0.62}{\degreeCelsius}, while the obtained oven temperature differs from our expertations (see section~\ref{sec:crystalselection}).

\subsection{UV light power output}

\begin{figure*}
	\includegraphics[width=8.5cm]{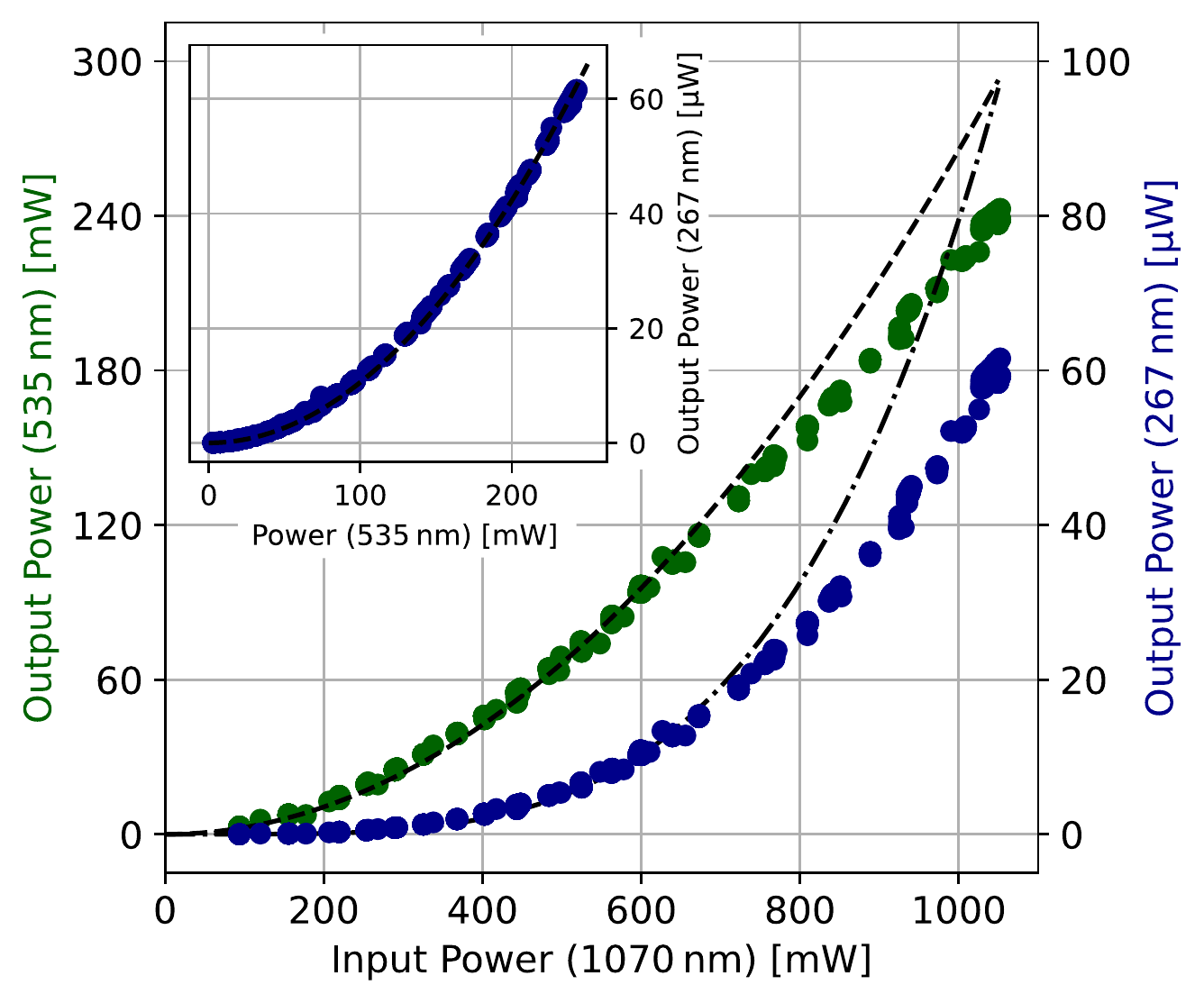}
	\caption{Green light power output at \SI{534.8}{\nano\meter} (green dots) generated by the first SHG and UV light power output at \SI{267.4}{\nano\meter} (blue dots) generated by the second SHG as a function of the IR pump light power at \SI{1069.6}{\nano\meter}. The dashed line shows a quadratic fit of the green light power with an output power of less than \SI{120}{\milli\watt}. The dashed-dotted line shows a fit with a fourth power function of the UV light power taking the same data selection into account. The inset shows the UV light power output at \SI{267.4}{\nano\meter} (blue dots) as a function of the Green light power as the pump light of the second SHG stage at \SI{534.8}{\nano\meter}. The dashed line is a quadratic fit of to the data.
	}
	\label{fig:3}
\end{figure*}

The optical power of the converted light from the first SHG at \SI{534.8}{\nano\m} and from the second SHG at \SI{267.4}{\nano\m} is measured as a function of pump light power at \SI{1069.6}{\nano\m}. The results are shown in Fig. \ref{fig:3}. It should be noted that the optimum oven temperature changes with pump light power as a consequence of additional crystal heating due to absorption. This is taken into account by optimizing the phase-matching temperatures of the two ovens for each pump power value individually. The IR pump light power of \SI{1}{\watt} is converted into more than \SI{200}{\milli\watt} green light by the first SHG and into more than \SI{50}{\micro\W} UV light power by the second SHG. For continuous operation of the FHG system, the green light output power should be limited to \SI{200}{\milli\watt} to avoid damage of the waveguide due to the photorefractive effect and green light induced infrared absorption \cite{sun_green-induced_2014}. As shown in Section 3, the SHG light power scales quadratically with the pump light power in the low intensity regime. Therefore, the data in Fig. \ref{fig:3} are fitted with a quadratic function for the first SHG stage, taking only data with green light power output of less than \SI{120}{\milli\watt} into account. We found a conversion coefficient of $\kappa=$\SI{119}{\percent\per\watt}, which is close to the conversion coefficient of $\kappa=$\SI{125}{\percent\per\watt} stated by the manufacturer. For higher power, pump beam depletion results in a linear scaling. For the FHG generation, the data is fitted to a function of fourth power, taking the same data as before into account. Only considering the second SHG stage, the UV light power output is a quadratic function of the green pump light as shown in the inset of Fig. \ref{fig:3}. Due to the low pump light power all data agree well with the quadratic fit. The conversion coefficient is $\kappa=$\SI{0.12}{\percent\per\watt}, which is close to the theoretical maximal value of $\kappa=$\SI{0.15}{\percent\per\watt}. Additionally, the output UV light power stability is measured for IR pump light power of \SI{800}{\milli\W} over \SI{10}{\hour}. The measured UV output power of approximately \SI{30}{\micro\W} is stable within a range of \SI{1}{\micro\W}, mainly limited by the power stability of the pump light laser. 

\section{UV light frequency stability}
\label{UVlref}
\begin{figure*}
	\includegraphics[width=12cm]{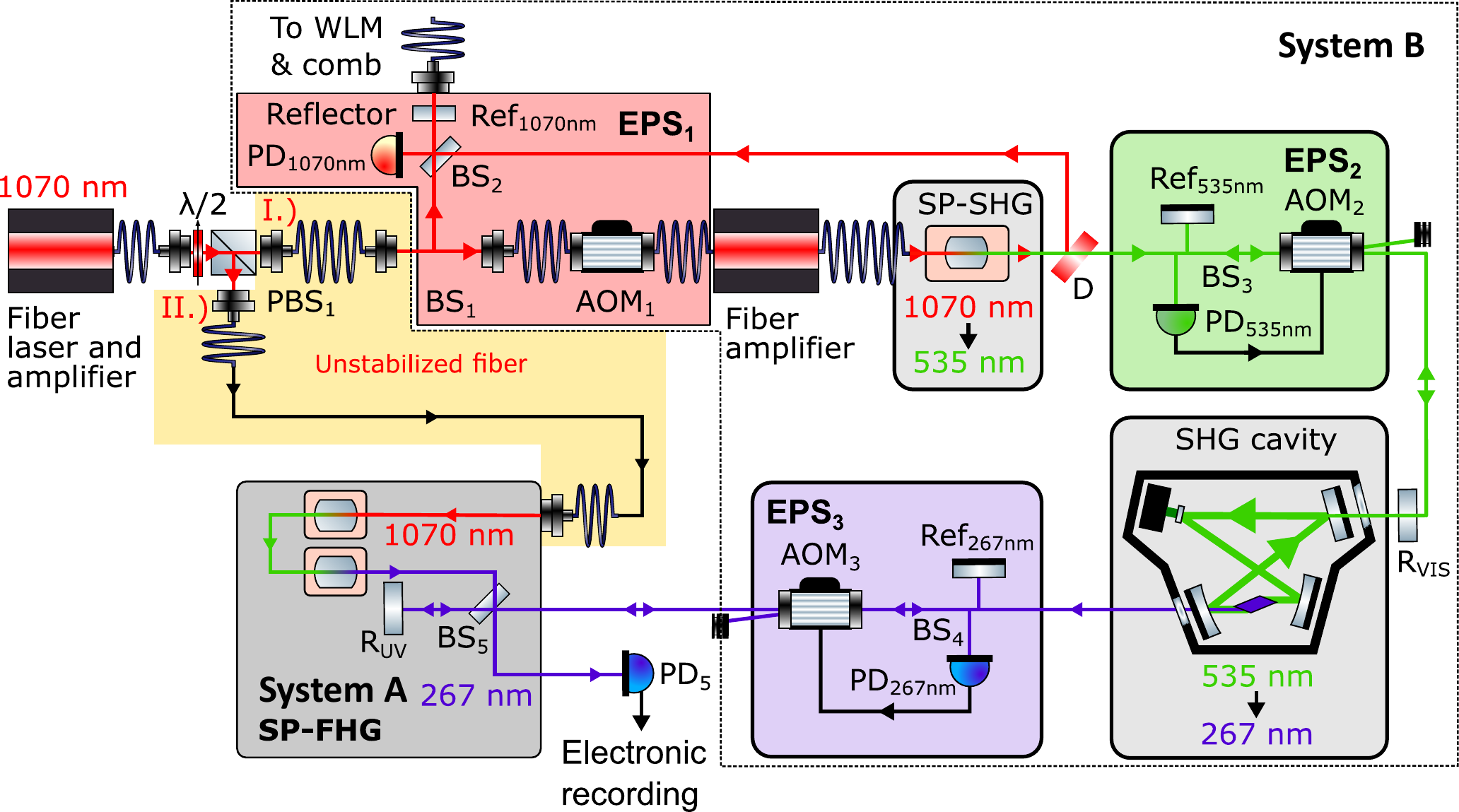} \caption{Simplified scheme for phase stability measurement with a second fourth harmonic generation system. 
	}
	\label{fig:FHG_comparison_setup}
\end{figure*}

The system described is designed for high passive frequency stability and at the same time offers the possibility of active stabilisation. In the following we refer to this system as system \emph{A}. For characterising the passive phase stability a beat-note measurement with an existing quadrupling system \emph{B} is performed, see Fig. \ref{fig:FHG_comparison_setup}. In this measurement, both laser signals are generated by the same fibre laser source. Therefore noise from the laser source itself is suppressed, since it is common to both arms. However, vibrations and temperature fluctuations can introduce optical path length variations in the reference system, which can significantly diminish the measured frequency stability and must be compensated for.

\subsection{Setup for frequency stability measurement}
\label{4a}
For the measurement of the phase stability variant 2 in Fig. \ref{drawing} is used. Here the outcoupling collimator C\textsubscript{2} is removed and the UV light generated by system \emph{B} enters the box through this port. The dichromatic mirror D\textsubscript{1} is replaced by a partially reflective window, which is used to overlap part of the light from both systems. However, most of the UV light from system \emph{B} is transmitted by the glass plate and is backreflected by D\textsubscript{1}, which serves as the reference mirror for the last part of the phase stabilisation scheme of system \emph{B}. The overlapping UV light inside the box is separated by the dichromatic mirror D\textsubscript{3} from the residual IR and green light and the beat-note of the UV light between system \emph{A} and \emph{B} is measured by the photodiode PD\textsubscript{5}.
\subsection{The reference UV laser source}
The laser source for both systems is a fiber laser oscillator with an instantaneous linewidth of $10\,$kHz whose output is amplified by a fiber amplifier. The common path between both setups ends at the polarising beamsplitter PBS\textsubscript{1} of Fig.~\ref{fig:FHG_comparison_setup} and some of the power is sent to system \emph{A} (single-pass FHG). About \SI{600}{\mW} are coupled into reference system \emph{B} through a \SI{5}{\m} long unstabilised fiber. An additional fiber amplifier in this system requires a special phase stabilisation scheme, since the amplifier can not be operated bidirectional. Therefore, the light transmitted through BS\textsubscript{1}, AOM\textsubscript{1}, the fiber amplifier and the fiber-coupled waveguide doubler (SP-SHG) must be phase-stabilized. For this, the IR reflection off dichroic mirror D is interfered with the reflection from mirror Ref\textsubscript{1070nm}, which acts as the reference mirror for phase stabilisation of the \SI{1069.6}{nm} light. By keeping the optical path between the fiber output, D and Ref\textsubscript{1070nm} at a minimal length and enclosed in a compact optical bench system, the phase at the reference point at Ref\textsubscript{1070nm} coincides with the phase of the fiber output and is transferred to the green light at dichroic mirror D. The paths from Ref\textsubscript{1070nm} and D interfere on BS\textsubscript{2} and the beat-note is measured by PD\textsubscript{1070nm}. A tracking oscillator filters the beat-note, which is phase-locked to a stable reference frequency by driving AOM\textsubscript{1} via a tunable direct digital synthesis (DDS) rf source. A second output of the tracking oscillator is counted using a dead-time free frequency counter.\\ 
The second doubling stage of system \emph{B} is based on a pump-beam resonant SHG cavity. Since multi-path interference prevents stabilisation of the green light through the cavity, optical paths before and after the SHG cavity are separately length stabilised, paying attention to keep the unstabilised optical paths as short as possible. \\
About \SI{172}{mW} at \SI{534.8}{\nm} of the waveguide's output is transmitted by the dichroic mirror D and sent through free-space AOM\textsubscript{2} into a  pump-beam resonant bow-tie cavity, where about \SI{4}{\mW} at \SI{267}{\nm} is generated in a BBO crystal. The length of the cavity is controlled by feedback onto a piezo mirror inside the cavity, using light at \SI{535}{\nm} reflected from the cavity's input window in a Hänsch-Couillaud locking scheme \cite{hansch_laser_1980}. Since multi-path interference prevents stabilisation of the green light through the cavity, optical paths before and after the SHG cavity are separately length stabilised, paying attention to keep the unstabilised optical paths as short as possible. In front of the cavity a small fraction of green light is back reflected onto PD\textsubscript{\SI{534.8}{nm}}, closing the path length stabilisation loop between reference mirror and retro-reflector by steering the frequency of AOM\textsubscript{2}.
The UV light generated in the cavity is frequency shifted by AOM\textsubscript{3}, then enters system \emph{A}. Here a fraction of the light is backreflected and brought into interference with the light from the reference arm.\\
At the point of PD\textsubscript{5} the light from system \emph{B} has accumulated a frequency offset of $f_B-f_A=\SI{251}{\MHz}$ with respect to \emph{A}. The beat signal from the photodiode is amplified and filtered by an tracking oscillator and counted by a dead-time free frequency counter. 

\subsection{Results of UV light frequency stability measurement}

\begin{figure*}
	\includegraphics[width=13cm]{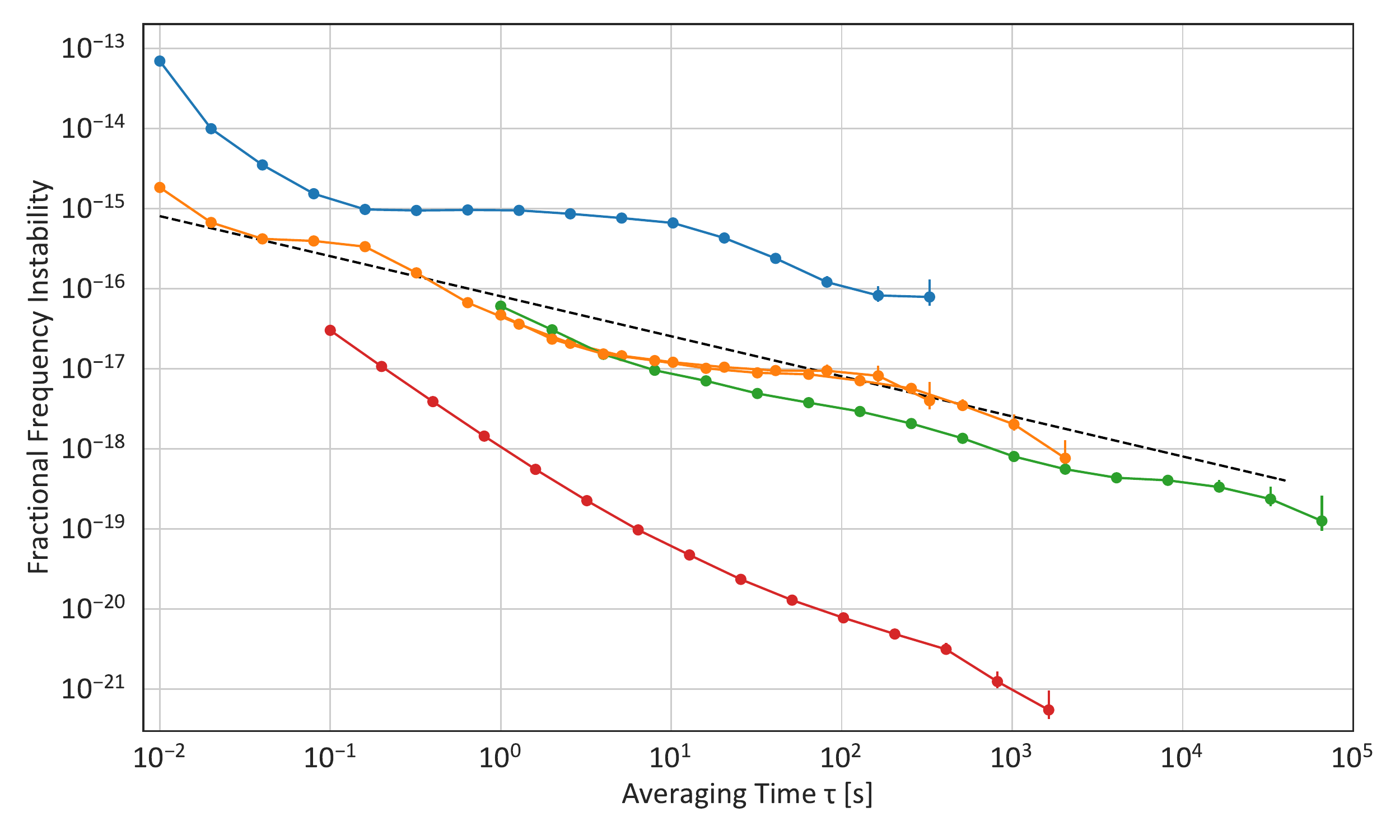}
	\caption{Measurements of the fractional frequency instability of the FHG system expressed as the modified Allan deviation. The blue (yellow) line results from a beat-note measurement with the unstabilised (stabilised) external UV source (system B), representing an upper fractional frequency instability limit for the single-pass FHG system. The green line shows the free-running signal of the interferometric phase stabilisation scheme divided by two without active phase stabilisation. A lower bound is given by electronic measurement noise of the in-loop error signal of the active phase stabilisation (red line). For comparison, the lifetime-limited quantum projection noise limit of the clock transition of \al at \SI{267.4}{\nm} is given as the dashed black line (see text for details).} 
	\label{fig.4}
\end{figure*}

To determine an upper bound for the frequency stability of the UV light the beat signal between the two systems is recorded for several hours, while system \emph{B} is either phase stabilised or unstabilised. System A (the single-pass FHG) is not phase stabilised during the measurements. Fig. \ref{fig.4} shows the modified Allan deviation\cite{benkler_relation_2015} of the recorded fractional frequency difference. For an averaging time of \SI{1}{\s} a fractional frequency instability limit of \SI{1e-15} is measured for the unstabilised systems (blue line). With reference system \emph{B} stabilised (yellow line) the fractional frequency instability limit reaches  \SI{5e-17} at a averaging time of \SI{1}{\s}, reaching down to \SI{e-18} for gate times of \SI{e3}{\s}. 

\subsection{In-Loop phase noise measurement of the single-pass FHG} 
For a further assessment of the intrinsic phase stability of the FHG, the interferometric phase stabilisation scheme as discussed in section~\ref{sec2} and shown in Fig.~\ref{drawing}, variant~1 is set up. The free-running signal of this interferometric phase stabilisation scheme is counted without active phase stabilisation. As the interferometric path contains any phase shift twice, we divide the result by two. From the data the modified Allan deviation of the fractional frequency instability is calculated. The result as shown in Fig. \ref{fig.4} (green line) gives the fractional frequency instability of the passive stability of the FHG system. At an integration time of \SI{1}{\s} we find for the fractional frequency instability \SI{7e-17}, which is in good agreement with the previous measurement using the stabilised external UV light source (system B) as a reference. For longer averaging times the fractional frequency stability of the free-running signal is lower compared to the measured fractional frequency instability measured with the reference system. This could either indicate a limitation of the phase stability of the reference system or is caused by different lab conditions during the measurements. Also remaining non-stabilised fibers in both systems in Fig.~\ref{fig:FHG_comparison_setup}) are likely to limit the measurement at the observed level. As the FHG setup is designed to be used to probe the clock transition of an \al ion, we compare our results to its quantum projection noise limit \cite{peik_laser_2006} using Ramsey spectroscopy with a probe time corresponding to the excited state lifetime of \SI{20.6}{\second} \cite{rosenband_observation_2007} (dashed black line in Fig.~\ref{fig.4}). 
The fractional frequency instability of the passive FHG system is already below the quantum projection noise limit. We expect further improvements of the phase stability of the FHG setup with active phase stabilisation. The measurement of the residual in-loop signal of the active phase stabilisation scheme (red) provides a lower bound to the achievable fractional frequency instability. The signal to noise ratio was below 40\,dB. For an independent verification of the performance of the stabilised system, two independent systems at this performance level would be necessary.\\ 
 \section{Conclusion}\label{conclusions}  
 The demonstrated single-pass FHG system has demonstrated exceptional phase stability while the provided output power is expected to be sufficient for probing clock transitions. In case more power is required while maintaining phase stability, the single-pass FHG system can act as a phase reference for a separate and more efficient enhancement cavity-based FHG system. An upper bound to the passive phase stability of the unstabilised setup is obtained by a frequency comparison with a stabilised external UV light source and by observing the free-running signal of the interferometric phase stabilisation loop. Both measurements show a phase stability close to the lifetime-limited quantum projection noise limit of an aluminium ion clock with a \SI{8}{\milli\hertz} narrow clock transition. Active phase stabilisation has been implemented and is expected to decrease the fractional frequency instability of the system further. 

\begin{backmatter}
\bmsection{Funding}
Funded by the Deutsche Forschungsgemeinschaft (DFG, German Research Foundation) – Project-ID 434617780 – SFB 1464, and under Germany’s Excellence Strategy – EXC-2123 QuantumFrontiers – 390837967. This project also received funding from the European Metrology Programme for Innovation and Research (EMPIR) coﬁnanced by the Participating 5 States and from the European Union’s Horizon 2020 research and innovation programme (Project No. 20FUN01 TSCAC). This project has been funded by the State of Lower Saxony, Hanover, Germany, through Niedersächsisches Vorab.

\bmsection{Disclosures}
The authors declare no conflicts of interest.

\bmsection{Data availability} Data underlying the results presented in this paper are not publicly available at this time but may be obtained from the authors upon reasonable request.

\end{backmatter}

\bibliography{paper_FHG.bib}

\end{document}